\documentclass[3p,twocolumn]{elsarticle}

\usepackage{lineno,hyperref,amsmath,amssymb}
\modulolinenumbers[5]

\journal{Physics Letters B}

\usepackage{graphicx}
\usepackage{dcolumn}
\usepackage{bm}
\usepackage{hyperref}
\hypersetup{colorlinks=true, citecolor=blue, urlcolor=blue, linkcolor=blue}
\usepackage{color}
\usepackage{epsfig}
\usepackage{epstopdf}
\usepackage{amsmath}
\def\nuc#1#2{\relax\ifmmode{}^{#1}{\protect\text{#2}}\else${}^{#1}$#2\fi}
\definecolor{bobcatgreen}{rgb}{0.3,0.6,0.1}

\begin{document}
  \newcommand {\nc} {\newcommand}
  \nc {\beq} {\begin{eqnarray}}
  \nc {\eeq} {\nonumber \end{eqnarray}}
  \nc {\eeqn}[1] {\label {#1} \end{eqnarray}}
  \nc {\eol} {\nonumber \\}
  \nc {\eoln}[1] {\label {#1} \\}
  \nc {\ve} [1] {\mbox{\boldmath $#1$}}
  \nc {\ves} [1] {\mbox{\boldmath ${\scriptstyle #1}$}}
  \nc {\mrm} [1] {\mathrm{#1}}
  \nc {\half} {\mbox{$\frac{1}{2}$}}
  \nc {\thal} {\mbox{$\frac{3}{2}$}}
  \nc {\fial} {\mbox{$\frac{5}{2}$}}
  \nc {\la} {\mbox{$\langle$}}
  \nc {\ra} {\mbox{$\rangle$}}
  \nc {\etal} {\emph{et al.}}
  \nc {\eq} [1] {(\ref{#1})}
  \nc {\Eq} [1] {Eq.~(\ref{#1})}
  \nc {\Refc} [2] {Refs.~\cite[#1]{#2}}
  \nc {\Sec} [1] {Sec.~\ref{#1}}
  \nc {\chap} [1] {Chapter~\ref{#1}}
  \nc {\anx} [1] {Appendix~\ref{#1}}
  \nc {\tbl} [1] {Table~\ref{#1}}
  \nc {\Fig} [1] {Fig.~\ref{#1}}
  \nc {\ex} [1] {$^{#1}$}
  \nc {\Sch} {Schr\"odinger }
  \nc {\flim} [2] {\mathop{\longrightarrow}\limits_{{#1}\rightarrow{#2}}}
  \nc {\IR} [1]{\textcolor{red}{#1}}
  \nc {\IB} [1]{\textcolor{blue}{#1}}
  \nc{\IG}[1]{\textcolor{bobcatgreen}{#1}}
  \nc{\pderiv}[2]{\cfrac{\partial #1}{\partial #2}}
  \nc{\deriv}[2]{\cfrac{d#1}{d#2}}

\begin{frontmatter}

\title{Simulating core excitation in breakup reactions of halo nuclei using an effective three-body force}

\author[jgu,ulb]{P.~Capel\corref{cor1}}
\ead{pcapel@uni-mainz.de}
\author[ou,tud,emmi]{D.~R. Phillips}
\ead{phillid1@ohio.edu}
\author[tud,emmi]{H.-W. Hammer}
\ead{hans-werner.hammer@physik.tu-darmstadt.de}

\cortext[cor1]{Corresponding author}
\address[jgu]{Institut f\"ur Kernphysik, 
Johannes Gutenberg-Universit\"at Mainz,
Johann-Joachim-Becher Weg 45, 
D-55099 Mainz, Germany}
\address[ulb]{Physique Nucl\'eaire et Physique Quantique (C.P.~229)\\
Universit\'e libre de Bruxelles (ULB),
50 avenue F.D. Roosevelt, B-1050 Brussels, Belgium}
\address[ou]{Institute of Nuclear and Particle Physics and Department of Physics and Astronomy, Ohio University, Athens, OH 45701,USA}
\address[tud]{Technische Universit\"at Darmstadt, Department of Physics, 64289 Darmstadt, Germany}
\address[emmi]{ExtreMe Matter Institute EMMI, GSI Helmholtzzentrum f\"ur Schwerionenforschung GmbH, 64291 Darmstadt, Germany}

\begin{abstract}
  We extend our previous calculation of the breakup of ${}^{11}$Be
 using Halo Effective Field Theory  and the Dynamical Eikonal Approximation
  to include an effective ${}^{10}$Be-n-target force.
The force is constructed to account for the virtual excitation of ${}^{10}$Be to its low-lying $2^+$ excited state. In the case of breakup on a ${}^{12}$C target this improves the description of the neutron-energy and angular spectra, especially in the vicinity of the ${}^{11}$Be $\frac{5}{2}^+$ state. By fine-tuning the range parameters of the three-body force, a reasonable description of data in the region of the $\frac{3}{2}^+$ ${}^{11}$Be state can also be obtained. This sensitivity to the three-body force's range results from the structure of the overlap integral that governs the ${}^{11}$Be $s$-to-$d$-state transitions which it induces. 
 \end{abstract}


\end{frontmatter}

\nolinenumbers

\section{Introduction}

Since their discovery in the mid-80s halo nuclei have been the subject of intense experimental and theoretical study \cite{Tan96,Riisager:2012it}.
These nuclei, located on the edge of the valley of stability, are much larger than their isobars. Their unusual size can be seen as a manifestation of quantum-tunneling: one or two
loosely bound valence nucleons have a high probability to reside in the classically forbidden region outside the nuclear mean-field potential. The nucleus can thus be described as an extended, diffuse halo surrounding a compact core.
Archetypes are $^{11}$Be, a one-neutron halo, and $^{11}$Li, with two neutrons in its halo.

The case of $^{11}$Be is especially interesting because it has recently been computed \emph{ab initio} within the No-Core Shell Model with Continuum (NCSMC)~\cite{CNR16}
using the N$^2$LO$_{\rm sat}$~\cite{Ekstrom:2015rta}  Chiral Effective Field Theory ($\chi$EFT)  nucleon-nucleon interaction.
Moreover ${}^{11}$Be has received much experimental attention, with its breakup on both lead and carbon targets measured at GSI and RIKEN \cite{Palit:2003av,Fuk04}.
In this work, we focus on the latter experiment, and more particularly on the dissociation of $^{11}$Be on $^{12}$C at 67~MeV/nucleon \cite{Fuk04}.

The presence of a halo in $^{11}$Be implies that the valence neutron strongly decouples from the other nucleons and hence that the structure of the nucleus can be described within a two-cluster model: a neutron loosely bound to a $^{10}$Be core.
Because of this clear separation of scales ${}^{11}$Be is well suited for the application of EFT~\cite{Bertulani:2002sz,Bedaque:2003wa,Hammer:2011ye}.
In this ``Halo EFT", the Hamiltonian that describes the core-halo structure is expanded as a series in a small parameter that is the ratio of the nucleus' small core radius to its large halo radius. Since the EFT is designed to be insensitive to short-distance details each term in the expansion of the core-halo interaction is taken to be a contact term or derivatives thereof. 
The parameters of this expansion, viz. the coefficients of each term in the core-neutron potential, are constrained by information on the structure of the nucleus, taken from experiment or from reliable nuclear-structure calculations.
Therefore, at each new order, more information is provided about the structure of ${}^{11}$Be
 in a systematic manner. In this way the key degrees of freedom can clearly be identified and ranked in importance (see Ref.~\cite{Hammer:2017tjm} for a recent review).

In a previous work \cite{CPH18}, we coupled such a Halo EFT description of $^{11}$Be to a precise reaction model in order to analyse the breakup data of Ref.~\cite{Fuk04}.
The reaction was described in the Dynamical Eikonal Approximation (DEA), which provides reliable collision observables in these experimental conditions \cite{BCG05,GBC06}.
Excellent agreement with the data of Ref.~\cite{Fuk04} on a Pb target was obtained. 
This idea has then been successfully extended to analyse the GSI experiment \cite{MC19}, transfer \cite{YC18} and knockout reactions \cite{HC21}.

For the $^{12}$C target, the magnitude and general shape of the breakup cross section of $^{11}$Be was well reproduced, but the calculation missed breakup strength in the energy region of the $\fial^+$ and $\thal^+$ resonances of $^{11}$Be at, respectively, 1.27~MeV and 3~MeV above the one-neutron threshold \cite{CPH18,MC19}.
The experimental breakup cross section exhibits clear peaks at these energies~\cite{Fuk04,CGB04}. 
We adjusted the $^{10}$Be-n interaction in the $d_{5/2}$ and $d_{3/2}$ partial waves to reproduce these continuum states as single-particle resonances.
This reduced the discrepancy between our prediction and data, although a significant breakup strength was still missing, especially at the $\thal^+$ resonance \cite{CPH18}.
These results were insensitive to off-shell properties of the ${}^{10}$Be-neutron interaction.

In the present work, we explore the significance of the core excitation in the breakup of $^{11}$Be on $^{12}$C by introducing in the reaction model a three-body interaction between the target, the $^{10}$Be core and the halo neutron.
This has the effect of inducing additional $s$-to-$d$-wave transitions in the ${}^{11}$Be system. 
Such an effective way to describe the virtual excitation of one of the participants in a collision has been used in various nuclear-physics contexts, from the $\Delta(1232)$ in the original Fujita-Miyazawa three-nucleon force
  \cite{Fujita:1957zz}, to proton scattering from deformed nuclei~\cite{Love:1967,Satchler:1971}, to $\chi$EFT nuclear forces~\cite{Ordonez:1993tn,vanKolck:1994yi,Epelbaum:2007sq} and transfer reactions \cite{Din19}. 
  For a colloquium on effective three-body forces in nuclear physics and
  beyond, see Ref.~\cite{Hammer:2012id}.
Here we expect that such an approach will be useful for energies less than the 3.368 MeV needed to produce the $2^+$ excitation of ${}^{10}$Be in the final state of the collision.
Therefore we mainly focus on relative energies $E$ between the $^{10}$Be core and the
 halo neutron below 3~MeV, where the Halo EFT expansion is well
 justified.
 However, our description can be extended to higher energies by tuning the range parameters of the three-body interactions.

After a brief reminder of the theoretical framework, we introduce the form of the three-body force that we consider in this work, see \Sec{formalism}.
Our results and their analysis are provided in \Sec{results}. We offer our conclusions in \Sec{conclusion}.

\section{Formalism}\label{formalism}

To model the collision of the one-neutron halo nucleus $^{11}$Be on a target, we describe it as a valence neutron n loosely bound to a $^{10}$Be core $c$~\cite{BC12}.
This two-body structure is modelled by the single-particle Hamiltonian
\beq
H_0=-\frac{\hbar^2}{2\mu}\Delta+V_{c\rm n}(\ve{r}),
\eeqn{e1}
where $\mu$ is the $c$-n reduced mass, $\ve{r}$ their relative coordinate, and $\Delta$ the corresponding Laplacian.
As in our previous work \cite{CPH18}, the binding potential $V_{c\rm n}$ is built within the framework of Halo EFT \cite{Hammer:2017tjm}. In this approach the details of $V_{c \rm n}$ are not important: their impact on observables is suppressed because of the broad extension of the halo wave function.
Accordingly, $V_{c\rm n}$ is taken to be a contact interaction and its derivatives.
For practical use, that interaction is regularised by a Gaussian of range $r_0$, and is parameterised partial wave by partial wave.
At next-to-leading order (NLO) we follow Ref.~\cite{CPH18} and take the potential in both the $s_{1/2}$ and $p_{1/2}$ channels to be:
\beq
V_{c\rm n}(r)=V_0(r_0) e^{-\frac{r^2}{2r_0^2}}+V_2(r_0) r^2  e^{-\frac{r^2}{2r_0^2}}.
\eeqn{e2}

The internal structure of the projectile is then described by the eigenstates of $H_0$.
The negative-energy states are discrete and correspond to the projectile bound states, whereas the positive-energy states form the continuum that simulates the broken-up ${}^{11}$Be.
As explained in Ref.~\cite{CPH18}, the potential depths $V_0$ and $V_2$ are fitted to reproduce structure information about the nucleus. 
In the present case we take the one-neutron separation energy of $^{11}$Be in both its bound states from experiment and the corresponding asymptotic normalisation coefficients (ANC) from the \emph{ab initio} NCSMC calculations of Calci \etal\ \cite{CNR16}.
 
Because our goal is to study the influence of the excitation of the core on the resonant breakup mechanism, we consider the potential we developed in Sec.~VII of Ref.~\cite{CPH18} and called \emph{beyond NLO}. I.e., in addition to fitting the potential \eq{e2} in the $s_{1/2}$ and $p_{1/2}$ waves, we also add a $c$-n interaction in both the $d_{5/2}$ and the $d_{3/2}$ partial waves and adjust it to produce a single-particle resonant state at the energy and with the width deduced from experiments.
We use the potential of range $r_0=1.2$~fm, apart from the $p_{3/2}$ partial wave, where we use the potential with $r_0=1.0$~fm that better reproduces the $p_{3/2}$ phaseshift predicted in Ref.~\cite{CNR16}.

The interaction between the projectile's constituents and the target $T$ is simulated by the optical potentials $V_{cT}$ and $V_{{\rm n}T}$. Their imaginary parts account for the channels not explicitly included in this few-body model of the collision, such as the capture of the halo neutron by the target or the dissociation of the core during the collision (see Ref.~\cite{CPH18} for details).

Within this framework, studying the projectile-target ($P$-$T$) collision amounts to solving the \Sch equation with the three-body Hamiltonian \cite{BC12}
\beq
H&=&-\frac{\hbar^2}{2\mu_{PT}}\Delta_R+H_0+V_{cT}(R_{cT})\nonumber \\
&& + V_{{\rm n}T}(R_{{\rm n}T}) +V_{3b}(\ve{R}_{cT},\ve{r}),
\eeqn{e3}
where $\mu_{PT}$ is the $P$-$T$ reduced mass, $\ve{R}$ their relative coordinate, and
  $\ve{R}_{cT}$ and $\ve{R}_{{\rm n}T}$ the $c$-$T$ and n-$T$ relative coordinates, respectively.
The \Sch equation has to be solved with the initial condition that the projectile in its ground state is impinging on the target.
At the intermediate beam energy considered here, this is a problem for which the Dynamical Eikonal Approximation (DEA) \cite{BCG05,GBC06} is perfectly suited. 

\begin{figure}
  \begin{center}
    \includegraphics[width=7.5cm]{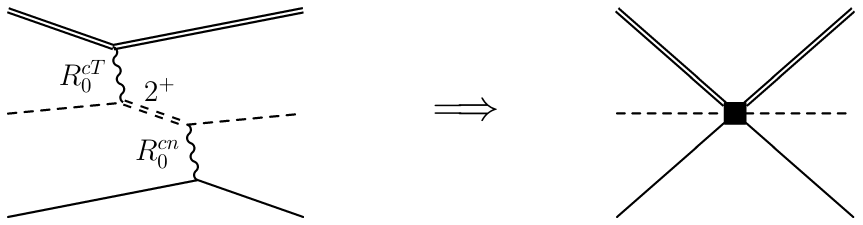}
\end{center}
  \caption{\label{f1}Illustration of the effective three-body force that
      arises from the virtual excitation of the $^{10}$Be core to
      the $2^+$ state. At long wavelengths/low energies the mechanism shown in the left part of the figure can be replaced by the effective three-body force shown in the right part. $R_0^{cT}$ and $R_0^{c\rm n}$ represent, respectively, the range of the core-target and core-neutron interactions.}
\end{figure}

The term $V_{3b}$ will always be present in an effective description of the collision at low resolution~\cite{Hammer:2012id}. It accounts for virtual excitations of the core or target to states not explicitly included in the Hilbert space. 
In the present case, this interaction is tailored to simulate the virtual excitation of $^{10}$Be during the collision as illustrated in \Fig{f1}.
The core, initially in its $0^+$ ground state (single dashed line), can be excited to its first excited $2^+$ state (double dashed line) through its interaction with the target (double solid line).
It can then come back to its ground state by interacting with the halo neutron (single solid line).
After tracing over the (unobserved) ${}^{10}$Be degrees of freedom this produces a quadrupole operator $(\nabla_{cT} \cdot \nabla_{c\rm n} \nabla_{cT} \cdot \nabla_{c\rm n} - \frac{1}{3} \nabla_{cT}^2 \nabla_{c\rm n}^2)$ acting on a product of Gaussian interactions: a core-target Gaussian potential with range $R_0^{cT}$ and a core-neutron one with range $R_0^{c\rm n}$. This yields:
\beq
\lefteqn{V_{3b}(\ve{R}_{cT},\ve{r}) = V_0^{3b}(R_0^{cT},R_0^{c\rm n}) \, Y_2^0\left(\widehat R_{cT}\cdot\widehat r,0\right)}\nonumber\\
\lefteqn{\quad\times \left(\frac{R_{cT}}{R_0^{cT}}\right)^2 e^{-\left(\frac{R_{cT}}{R_0^{cT}}\right)^2} \left(\frac{r}{R_0^{c\rm n}}\right)^2 e^{-\left(\frac{r}{R_0^{c\rm n}}\right)^2},}
\eeqn{e4}
where $ Y_l^m(\theta,\phi)$ is a spherical harmonic,
$V_0^{3b}(R_0^{cT},R_0^{c\rm n})$ is the magnitude of the three-body interaction for particular ranges $R_0^{cT}$ and $R_0^{c\rm n}$.
\footnote{We divide the force by these regulator scales $(R_0^{cT})^2$ and $(R_0^{c\rm n})^2$ so that $V_0^{3b}(R_0^{cT},R_0^{c\rm n})$ has dimensions of energy.}
The three-body force in Eq.~(\ref{e4}) is very similar to the effective force derived already fifty years ago by Love and Satchler to account for core polarization in low-energy proton scattering from deformed nuclei~\cite{Love:1967,Satchler:1971}. A quantitative comparison of that approach with ours is given in the next section.
Note that the inclusion of core degrees of
  freedom via the effective three-body force, Eq.~(\ref{e4}), does not change
  the two-body spectrum. As a consequence, the resonance
  energies of $^{11}$Be are not modified and the two-body force can be adjusted so that it reproduces the physical
  resonance energies.
  This is a clear advantage compared to other approaches that account for core
  excitation \cite{Vinh95,NTJ96,Tar06,BPB17}.

 At higher energies, the $2^+$ state can be excited as a real degree of freedom and this description using an effective three-body force
 breaks down. We now study the impact of $V_{3b}$ on ${}^{11}$Be breakup observables in the region $E \lesssim 3$~MeV and their sensitivity to these parameters.

\section{Results}\label{results}

\begin{figure*}
\center
\includegraphics[width=6.cm]{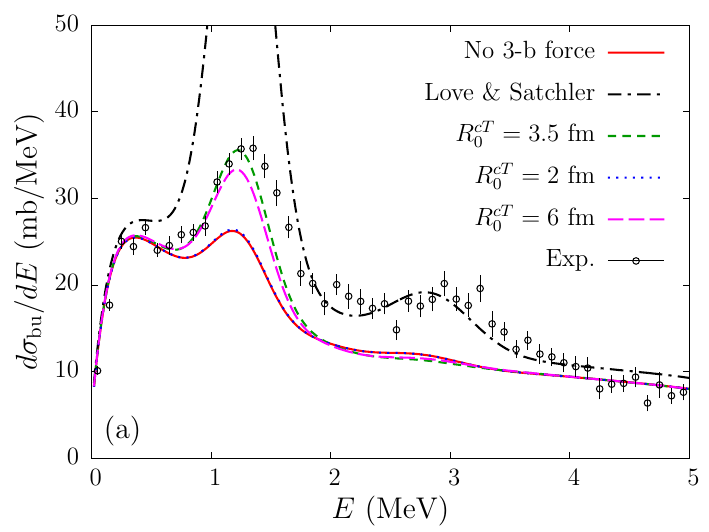}
\includegraphics[width=6.cm]{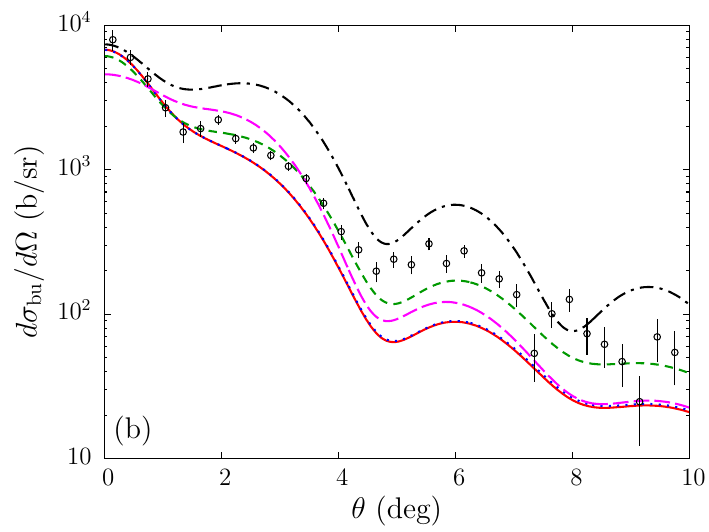}
\caption{\label{f2} Influence of a three-body force  on the breakup cross section for $^{11}$Be on $^{12}$C at 67~MeV/nucleon
(a) as a function of the $^{10}$Be-n relative energy $E$ after dissociation;
(b) as a function of the scattering angle $\theta$ of the $^{10}$Be-n centre of mass for $1.2~\mbox{MeV}\le E \le 1.4~\mbox{MeV}$, viz. at the $\fial^+$ resonant state.
The prediction of Love and Satchler \cite{Love:1967,Satchler:1971} is compared to the EFT approach \eq{e4} with different ranges $R_{0}^{cT}$ (we keep $R_0^{c\rm n}=1.7$~fm fixed by the range of $V_{c\rm n}$) and to the data of Ref.~\cite{Fuk04}.}
\end{figure*}

We look in particular at two observables measured at RIKEN \cite{Fuk04}.
The first is the breakup cross section expressed as a function of the relative energy $E$ between the $^{10}$Be core and the halo neutron after dissociation, ``the energy distribution", see \Fig{f2}(a).
Because we focus on the excitation of $^{11}$Be resonant states during the breakup, we also look at the angular distribution, in which the breakup cross section is computed as a function of the scattering angle $\theta$ of the $^{10}$Be-n centre of mass within the energy range of the $\fial^+$ resonance, $1.2~\mbox{MeV}\le E \le 1.4~\mbox{MeV}$, see \Fig{f2}(b).

Figure~\ref{f2} illustrates the results of our first series of tests.
So that we can compare to the experimental data of Ref.~\cite{Fuk04} the theory curves in the left panel have been folded with the experimental energy resolution (a Gaussian with $\sigma=0.191~{\rm MeV}^{1/2}\sqrt{E}$ \cite{Fuk04}); those in the right panel were folded with the angular resolution (a Gaussian with $\sigma=0.48^\circ$ \cite{Fuk04}).
In Fig.~\ref{f2}(b) we accounted for the finite experimental energy resolution by normalising the result from the angular folding to the integral of the energy distribution [Fig.~\ref{f2}(a)] in the vicinity of the $\fial^+$ resonance.
The red solid line represents the theoretical cross sections obtained without three-body force, as found in Ref.~\cite{CPH18}.
Although the description of $^{11}$Be includes resonant states in both the $d_{5/2}$ and $d_{3/2}$ waves, their effect in the calculation is far less than the peaks seen in the data at these resonance energies. 
The $d_{5/2}$ resonance leads to only half of the required strength to reproduce the data in the region of the $\fial^+$ state, and the effect of the $d_{3/2}$ resonance is barely visible after folding.
This result is unaffected by the value of $r_0$ that is chosen, i.e., it seems independent of the off-shell behavior of $V_{c\rm n}$. 
We concluded that our simple three-body model of the reaction lacked some significant degrees of freedom.
According to the work of Moro and Lay \cite{ML12}, one possibility is the excitation of the $^{10}$Be core to its $2^+$ excited state.

The Love and Satchler approach to including such core-polarization effects yields a  term to be added to the bare two-body interactions whose form is very similar to \Eq{e4}. But it also says that the three-body force's strength and range can be computed by taking the derivative of the two-body forces $V_{cT}$ and $V_{c\rm n}$ with respect to their ranges in the $R_{cT}$ and $r$ coordinates.
The corresponding results are displayed in \Fig{f2} as the black dash-dotted lines.
Although the agreement with the data at energies $E\gtrsim2$~MeV is very good, this correction over-predicts the
strength of the three-body force by a factor of five in the region of the $\fial^+$ resonance.
Love and Satchler's force thus has the correct general form, but does not provide the right magnitude.
This is not surprising for a near-dripline nucleus like $^{11}$Be.
In particular the mean-field approach assumed for the interaction between the valence nucleon and the core is not valid for $^{11}$Be, which exhibits a well-known shell inversion between its ground and first excited states.
Moreover, the strength of $V_{c\rm n}$ varies strongly with its range $r_0$ (see, e.g., Tables~I and II of Ref.~\cite{CPH18}).
The estimate of the derivative of that two-body force to calculate the strength $V_0^{3b}$ is therefore marred with uncertainty.

\begin{figure*}
\center
\includegraphics[width=6.cm]{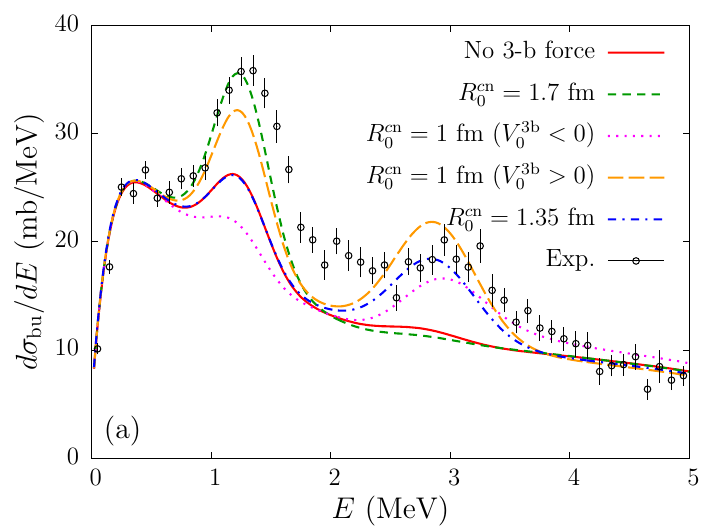}
\includegraphics[width=6.cm]{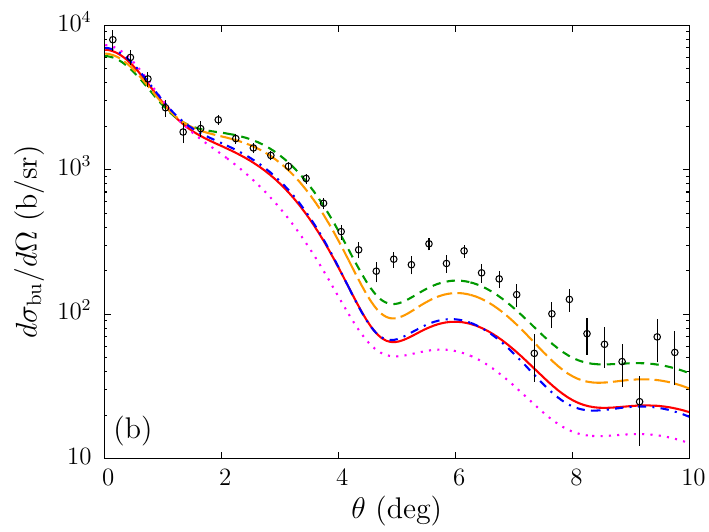}
\caption{\label{f3}Influence of the range of the three-body force \eq{e4} in the $c$-n relative coordinate $R_0^{c\rm n}$ in the breakup of $^{11}$Be on $^{12}$C at 67~MeV/nucleon.
(a) Energy distribution; (b) angular distribution for the breakup at energies around the $\fial^+$ resonance.
The range $R_0^{cT}$ is kept to its 3.5~fm optimum.
Experimental data from Ref.~\cite{Fuk04}.}
\end{figure*}

Our EFT approach is similar to that of Refs.~\cite{Love:1967,Satchler:1971}, since our three-body force exhibits the same form as the induced interaction derived in that work.
However, there are important differences too.
From the EFT perspective, core polarization is only one among several effects that can induce three-body forces.
Therefore we believe that  a better strategy than predicting the strength of the ${}^{10}$Be-neutron-${}^{12}$C three-body force
is to fit that strength to data.
We therefore add the three-body potential \eq{e4} to our model and run a series of DEA calculations that probe the parameter space of that force.
The results of these tests are illustrated in Figs.~\ref{f2} and \ref{f3}.

A first natural choice is to assume $R_0^{cT}=3.5$~fm and $R_0^{c\rm n}=\sqrt{2}r_0=1.7$~fm, which are the ranges of $V_{cT}$ and  $V_{c\rm n}$, respectively.
The choice $V_0^{3b}=-100$~MeV then leads to good agreement with the data in the $\fial^+$ peak, see the green short-dashed line in \Fig{f2}(a).
This good result is confirmed in the angular distribution shown in \Fig{f2}(b): the general shape of the experimental cross section is very well reproduced at nearly all angles.

The expression \eq{e4} involves a quadrupole excitation and therefore at first order it can take ${}^{11}$Be from its ground $s_{1/2}$ state to a $d$ wave in the continuum. Dynamical effects, such as couplings within the continuum, could mean that the three-body force also indirectly increases the breakup contribution of other partial waves.
However, our calculations show this is not the case: not only is the increase in the breakup strength limited to the $d$ waves, but it affects only the energy range of the $d$ resonances.

The effect of the three-body force in the $\thal^+$ peak is only marginal.
The breakup cross section in that region actually decreases slightly compared to the energy distribution obtained without a three-body force.
For this natural choice of ranges, we have not found a way to sufficiently populate that resonance while keeping the good agreement obtained for $1.2~\mbox{MeV}\le E \le 1.4~\mbox{MeV}$.
Near the $\thal^+$ peak, the resolution reaches the limits of an effective three-body force description of the $2^+$ excitation of the core.
This is in accord with the results of Ref.~\cite{ML12}, which found it necessary to include a configuration in which $^{10}$Be is in its $2^+$ excited state in the Hilbert space of the model in order to reproduce the data around $E=3$~MeV.

To test the sensitivity of our model to the values of $R_0^{cT}$ and $R_0^{c\rm n}$, we explore those parameters of the model space.
First we vary the $c$-$T$ range in \Eq{e4}.
Too small an $R_0^{cT}$, such as 2~fm shown in \Fig{f2} (blue dotted lines) leads to nearly unnoticeable effects at all energies and all angles, despite a significant magnitude ($V_0^{3b}=-500$~MeV).
Similar results are obtained with $R_0^{cT}=1$~fm.
When the three-body-force range is this small it acts only within the distance at which the interaction between the core and the target is dominated by the absorption channel.
The effect of any additional real interaction then vanishes.
For this three-body force to have the desired effect it must have a range that exceeds, or equals, that of the imaginary part of the $c$-$T$ optical potential.

Choosing larger values of $R_0^{cT}$ also leads to unsatisfactory results.
First, $R_0^{cT}\gg 3.5$~fm contradicts the idea of EFT, because it should correspond to the range of the short-distance physics neglected in the problem.
Second, it produces deleterious phenomenological consequences too: 
calculations performed with $R_0^{cT}=6$~fm (and $V_0^{3b}=-100$~MeV) are shown in magenta long-dashed lines in \Fig{f2}.
This $V_{3b}$ still produces good agreement with the experimental energy distribution in the $\fial^+$ resonance.
However, in the angular distribution, that agreement is reduced compared to $R_0^{cT}=3.5$~fm: the theoretical cross section does not exhibit the proper angular dependence at forward angle and it decays too rapidly with the scattering angle, leading to a clear underestimation of the data at larger angle.
Both these problems worsen when that range is extended to 8 or 10~fm.

Summarizing the story so far: a well-tailored three-body potential can provide the strength in the $\fial^+$ resonance that was missing from the earlier Halo EFT + DEA calculation with two-body potentials alone.
The range $R_0^{cT}$ should be chosen close to its ``natural'' value, i.e. the range of the optical potential $V_{cT}$.
As long as $3~{\rm fm} \leq R_0^{cT} \leq 5~{\rm fm}$ the magnitude of $V_{3b}$ can be chosen to reproduce the energy distribution up to about 1.5~MeV and the angular distribution integrated over $1.2~\mbox{MeV}\le E \le 1.4~\mbox{MeV}$.

Up to now, $R_0^{c\rm n}$ has been kept to 1.7~fm, fixed by the range $r_0$ of $V_{c\rm n}$.
In this second step of our analysis, it is varied while $R_0^{cT}$ is maintained to its optimal value 3.5~fm

The results are shown in \Fig{f3} for (a) the breakup energy distribution and (b) the angular distribution for the breakup towards the $\fial^+$ resonance.
The already discussed results without three-body force and with the three-body force with $R_0^{c\rm n}=1.7$~fm are shown as the red solid and green short-dashed lines.
Exploring the $R_0^{c\rm n}$ model space to smaller distances produces interesting results.
Choosing, e.g., $R_0^{c\rm n}=1$~fm while keeping an attractive three-body force ($V_0^{3b}=-1000$~MeV), leads to a significant $\thal^+$ peak and a reduction of the breakup cross section in the region of the $\fial^+$ continuum state (magenta dotted lines).
This suggests that simulating the core excitation by a three-body force can also lead to the excitation of the $\thal^+$ resonance as a single-particle state---as long as the $c$-n interaction range is chosen small enough.

\begin{figure}[htb]
\center
\includegraphics[width=6.cm]{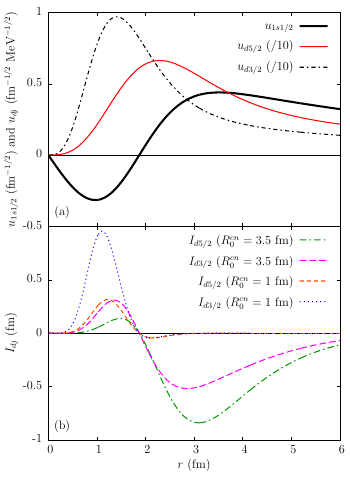}
\caption{\label{f4} (a) Wave functions of the states involved in the resonant breakup of $^{11}$Be. (b) The product of these wave functions with the $r$-dependence of the three-body force $I_{lj}$ defined in \Eq{e6}.
}
\end{figure}

To elucidate why the choice of the $c$-n interaction range has such a profound impact on the action of the three-body force on the cross section, we
display in \Fig{f4}(a) the wave functions obtained from the Halo EFT description of $^{11}$Be for the $1s_{1/2}$ ground-state (thick black solid line), and for the $d_{5/2}$ (red solid line) and $d_{3/2}$ (black dash-dotted line) resonant states (the wave functions of the resonant states are divided by ten for readability).
Both resonant wave functions exhibit similar features: at short distance they present a bound-state like peak before starting to oscillate at larger distances.
However, these peaks appear on the opposite sides of the node of the bound-state wave function located at $r\approx 2$~fm.
Whereas $u_{d5/2}$ peaks after the node, $u_{d3/2}$ exhibits its maximum at lower radius. 

This selectivity of the $d$ states according to $R_0^{c\rm n}$ value then becomes clear from \Fig{f4}(b) where we plot the overlap of the radial wave functions of the initial $1s_{1/2}$ bound state and the final $d$ resonant state, multiplied by the $r$-dependence of an attractive three-body force \eq{e4}
\beq
I_{dj}(r) = - u_{dj}(r) \left(\frac{r}{R_0^{c\rm n}}\right)^2 e^{-\left(\frac{r}{R_0^{c\rm n}}\right)^2} u_{1s_{1/2}}(r).
\eeqn{e6}
This integral would appear in a first-order description of the reaction, and it explains the effect of the three-body force seen in Figs.~\ref{f2} and \ref{f3}.
With $R_0^{c\rm n}=1.7$~fm, $V_{3b}$ excites the ${}^{11}$Be projectile at large $r$.
The dominant part of $I_{d5/2}$ (green dash-dotted line) is located beyond 2~fm, and is negative.
On the contrary, the major contribution to $I_{d3/2}$ (magenta long-dashed line) is located at short distances, and is positive.
While the former adds up to the effect of the (attractive) two-body forces, increasing the population of the $d_{5/2}$ resonance, the latter opposes $V_{cT}$ and $V_{{\rm n}T}$, hence reducing the breakup of $^{11}$Be towards its $d_{3/2}$ resonance.
Accordingly $\fial^+$ excitation is favored over $\thal^+$ excitation for $R_0^{c\rm n}=1.7$~fm  [see the green short-dashed lines in Figs.~\ref{f2}(a) and \ref{f3}(a)].
Similar results are obtained for $R_0^{c\rm n}$ between 2 and 4~fm, provided $V_0^{3b}$ is adjusted appropriately.

When the $c$-n range of the three-body interaction is reduced to $R_0^{c\rm n}=1$~fm, the influence of the large radii on $I_{lj}$ is reduced to a trickle: the major contribution comes now from $r\lesssim 2$~fm.
Here the $d_{3/2}$ wave function dominates [compare the blue dotted and orange short-dashed lines in \Fig{f4}(b)].
This short-range contribution is positive, so it opposes the breakup strength of the two-body forces.
However, because the breakup strength for the $d_{3/2}$ resonance generated by $V_{cT}$ and $V_{{\rm n}T}$ is small, the large effect of the three-body force observed here is sufficient to reproduce the experimental $\thal^+$ peak, see the magenta dotted line in \Fig{f3}(a).
This $^{11}$Be resonant state is therefore mostly populated through the excitation of the $^{10}$Be core.
At the $d_{5/2}$ resonance, the contribution of that three-body interaction cancels the effect of the two-body optical potentials.
This explains the decrease-increase in the $d_{5/2}$-$d_{3/2}$ peaks seen when comparing the magenta dotted and red solid lines in \Fig{f3}(a).
It also explains why the corresponding angular distribution at the $d_{5/2}$ resonance is overall suppressed, see the panel (b) of that figure. 

While the high energy of the $\thal^+$ resonance clearly stretches an EFT description without explicit $2^+$ core excitation, this result suggests a way to excite both the $\thal^+$ and $\fial^+$ resonances simultaneously with an effective three-body force: use a short-range repulsive force ($V_{0}^{3b}>0$).
We note that since the effective three-body force is not observable by itself and its strength $V_0^{3b}$ is resolution dependent, it is reasonable for $V_0^{3b}$ to change from negative to positive as $R_0^{c\rm n}$ is decreased~ \cite{Bedaque:1998kg,Hammer:2012id}.
The results of a calculation with $R_0^{c\rm n}=1$~fm and $V_0^{3b}=+1000$~MeV are displayed in \Fig{f3} as the orange long-dashed lines.
Now both resonances are excited and exhibit breakup strengths in qualitative agreement with the data.
This three-body interaction also provides an angular distribution at the $d_{5/2}$ resonance in excellent agreement with the data.
We have obtained similar results with $R_0^{c\rm n}=0.8$ and 1.1~fm so this result is not strongly sensitive to the choice of that parameter, as long as it is small.
A singular effect is observed using $R_0^{c\rm n}=1.35$~fm, see the blue dash-dotted curves in \Fig{f3}.
At that value, despite the presence of a significant three-body force in the calculation ($V_0^{3b}=1000$~MeV), we do not observe any significant change in the $d_{5/2}$ resonant breakup in either the energy or angular distribution compared to the case without three-body force.
Now the two lobes of $I_{d5/2}$ are close in magnitude but of opposite sign, leading to a near-exact cancellation of the effect of $V_{3b}$.

\section{Conclusion}\label{conclusion}

The nuclear breakup of $^{11}$Be on $^{12}$C excites the $\fial^+$ and $\thal^+$ resonances \cite{Fuk04}.
This reaction therefore constitutes an ideal tool to study these states above the one-neutron separation threshold \cite{Fuk04,CGB04}.
In a previous work \cite{CPH18}, we examined in detail the influence of the description of the projectile upon the reaction calculation, coupling a Halo-EFT description of $^{11}$Be \cite{Hammer:2017tjm} to the DEA \cite{BCG05,GBC06}.
Describing the $\fial^+$ and $\thal^+$ resonances as, respectively, a $d_{5/2}$ and $d_{3/2}$ neutron interacting with a $^{10}$Be core in its $0^+$ ground state is not sufficient to reproduce the peaks observed in the experimental cross section \cite{CPH18}.

In this Letter, we have presented an extension of Ref.~\cite{CPH18}, adding a three-body interaction between the $^{12}$C target, the $^{10}$Be core, and the halo neutron. This three-body force is tailored to simulate the effect of the virtual excitation of the ${}^{10}$Be core to a $2^+$ state during the reaction.
When the range of the three-body force in the core-target coordinate is of the order of that of the $c$-$T$ nuclear optical potential it is possible to find a realistic three-body-force strength that reproduces the experimental breakup cross section in the energy region of the $\fial^+$ resonant state without affecting the good agreement with the data at  $E<3$~MeV. 

A similar idea was previously used by Love and Satchler to 
  include core polarization effects in proton scattering from deformed
  nuclei~\cite{Love:1967,Satchler:1971}. They also gave a prescription
  to derive the strength of the force from the relevant optical potentials, and from the deformation of ${}^{10}$Be. However, applied to
  the nuclear breakup of $^{11}$Be this prescription overestimates the strength of the three-body force by roughly a factor of five.
   This observation clearly demonstrates the power of the EFT framework.
   The three-body force is not merely an efficient way to include the
   physics of a particular model, but allows to capture
   all the different physics mechanisms that can contribute to such a
   three-body force in a systematic way. In particular, it is able
   to accommodate the effects of the shell-inversion in $^{11}$Be.

More recently, Moro and Lay found that explicitly including the $2^+$ excited state of $^{10}$Be in the projectile description enables satisfactory reproduction of the experimental breakup cross section \cite{ML12}. 
In our complementary approach, we incorporate the core excitation in the effective operators that encode the reaction mechanism.
This ability to include virtual excitations in either operators or wave functions can be formalised through the use of unitary transformations~\cite{Okubo:1954zz,Lee:1980yrt} and suggests that it is difficult to gain model-independent insight into the structure of these resonances from reactions on a ${}^{12}$C target.
This is particularly true for the $\thal^+$ resonance, which exists close to the ${}^{10}$Be$(2^+)$-n threshold.
For this ${}^{11}$Be state a description that includes only virtual first-order excitation of the $2^+$ state may not be valid.

Tuning the range of the three-body interaction in the $c$-n coordinate results in the excitation of the $\fial^+$, the $\thal^+$, or both, resonances.  This selectivity can be traced back to the radial dependence of the overlap wave functions of the single-particle description of the states.
It highlights the fact that, in the region $E\gtrsim 3$~MeV, the resolution in the reaction reaches the limits of  an EFT description without an explicit $2^+$ degree of freedom, because details of the EFT's implementation can be resolved.

The results produced here confirm that Halo EFT~\cite{Hammer:2017tjm} is an efficient and flexible tool for reactions involving halo nuclei~\cite{CPH18,MC19,YC18,HC21,Schmidt:2018doj}.
In EFT the appearance of three-body forces is essential, since they account for the impact of missing degrees of freedom on observables~\cite{Hammer:2012id}.
In the case of ${}^{11}$Be, the reaction observables for which Halo EFT is accurate can be extended by using an EFT three-body force to account for the $2^+$ excitation of the ${}^{10}$Be core.
This enables us to explore the effect of virtual core excitation on breakup cross sections without resorting to a numerically expensive description of the projectile.
An obvious next step is to constrain this three-body interaction from structure or reaction inputs other than the ones we are trying to describe, or use it to describe other reactions as we have done in Refs.~\cite{CPH18,MC19,YC18,HC21} for Halo EFT, in order to fully exploit the universality of the low-energy constants of the EFT.
In the longer term a Halo EFT description of the projectile that explicitly includes the excitation of the core, similar to  that of Ref.~\cite{ML12}, would be an asset, as it would presumably extend the phenomenological reach of the calculation and shed further light on the structure of resonances in the ${}^{11}$Be system, especially its $\thal^+$ state. 

\section*{Acknowledgments}
This project has received funding from the European Union's Horizon 2020 research and innovation programme under grant agreement No 654002.
It was also supported by the Deutsche Forschungsgemeinschaft within the Collaborative Research Centers SFB 1044 (Projektnummer 204404729) and SFB 1245 (Projektnummer 279384907), by the Federal Ministry of Education and Research (BMBF) under
contract 05P18RDFN1, and the PRISMA (Precision Physics, Fundamental Interactions and Structure of Matter) Cluster of Excellence, the US Department of Energy (contract DE-FG02-93ER40756), and by the ExtreMe Matter Institute.  DRP is grateful for the warm hospitality of the IKP Theoriezentrum Darmstadt. 
PC acknowledges the support of the State of Rhineland-Palatinate. PC and DRP are grateful to the INT program INT 17-1a, ``Toward Predictive Theories of Nuclear Reactions Across the Isotopic Chart" which stimulated several of the ideas discussed in this manuscript.


\end{document}